\documentclass[12pt,a4paper]{article}
\usepackage{amssymb}
\usepackage{graphicx}
\newcounter{eg}
\newtheorem{eg}{Example}[section]   
\def\beg{\begin{eg}\rm}
\def\eeg{\hfill\sq\end{eg}}
\def\t#1{\tilde #1}
\def\ad{\mbox{ad}\,}

\def\c#1{{\cal #1}}

\def\Dirac{{\raise0.09em\hbox{/}}\kern-0.69em D}

\def\kbar{{\mathchar'26\mkern-9muk}}

\def\sq{\hbox{\rlap{$\sqcap$}$\sqcup$}}

\def\t#1{\tilde #1}
\def\tr{\mbox{Tr}}

\def\t#1{\tilde #1}
\def\ad{\mbox{ad}\,}

\def\c#1{{\cal #1}}

\def\Dirac{{\raise0.09em\hbox{/}}\kern-0.69em D}

\def\kbar {{\mathchar'26\mkern-9muk}}

\def\sq{\hbox{\rlap{$\sqcap$}$\sqcup$}}

\def\t#1{\tilde #1}
\def\tfrac #1#2{\textstyle{\frac{#1}{#2}}}

\def\tr{\mbox{Tr}\,} 

\def\k{\kern-.1em\mathbin{,}\kern-.1em}
\def\hk{\kern.12em\raise-1em\hbox{$\hat{\raise1em\hbox{,}}$}\kern.12em}
\newcommand{\initiate}{\setcounter{equation}{0}}
\begin{document}
\title{Can noncommutativity resolve the Big-Bang singularity?}

\author{M. Maceda$\strut^{1,a}$, \ J. Madore$\strut^{1,b}$ \\[10pt]
        P.~Manousselis$\strut^{2,3,c}$, \ G. Zoupanos$\strut^{4,d}$ \\[20pt]
        $\strut^{1}$Laboratoire de Physique Th\'eorique\\
        Universit\'e de Paris-Sud, B\^atiment 211, F-91405 Orsay
\and    $\strut^2$ Department of Engineering Sciences,\\
University of Patras, 26110 Patras, Greece \\
\and    $\strut^3$Physics Department, National Technical
University,\\
        Zografou Campus, GR-157~80 Zografou, Athens
\and    $\strut^4$ Theory Division, CERN, CH-1211, Geneva 23}

\date{}

\maketitle

\abstract{A possible way to resolve the singularities of general
  relativity is proposed based on the assumption that the description
  of space-time using commuting coordinates is not valid above a
  certain fundamental scale. Beyond that scale it is assumed that the
  space-time has noncommutative structure leading in turn to a
  resolution of the singularity. As a first attempt towards
  realizing the above programme
  a modification of the Kasner metric is
  constructed which is commutative only at
  large time scales. At small time scales, near the singularity, the
  commutation relations among the space coordinates
  diverge. We interpret this result as meaning that the singularity has been completely delocalized.}
\vspace{2.5cm}

\begin{flushleft}
CERN-TH Preprint 2003/048\\ $\strut^{a}$e-mail address:
maceda@th.u-psud.fr\\ $\strut^b$e-mail address:
John.Madore@th.u-psud.fr\\ $\strut^{c}$e-mail address:
pman@central.ntua.gr
\\ $\strut^{d}$ On leave from $3$. e-mail address: George.Zoupanos@cern.ch
\end{flushleft}

\eject

\parskip 4pt plus2pt minus2pt

\initiate
\section{Motivation}

It is folk wisdom that singularities and divergences are not
really physical but rather technical artifacts indicating the
limitations of the theory in which they appear. Most theories do
contain singularities as an essential element and these are
generally considered to contain important information on possible
extensions. Among elementary particle physicists for example it is
widely believed that infinities in renormalizable field theories
not only are not problematic since they can be treated by the
renormalization procedure but instead can be considered as signals
of a nearby new-physics threshold.  A possible way to resolve
field-theory singularities, that is, infinities related to the
field-theoretical description of particle physics is to introduce
still another scale this time related to the possible unification
scale of the non-gravitational interactions. Grand Unified
Theories (GUTs) with ${\cal N}=1$ supersymmetry have been
constructed which can be made finite even to all-loops, including
the soft supersymmetry breaking
sector~\cite{KapMonZou93,LucZou97,KobKubMonZou01,LucZou96}. There
exist a method to construct GUTs with reduced independent
parameters which consists of searching for renormalization-group
invariant (RGI) relations valid below the Planck scale and which
in turn are preserved down to the GUT
scale~\cite{LucZou96,KobKubMonZou98}. Of particular interest is
the possibility of finding RGI relations among couplings which
guarantee finiteness to all-orders in perturbation theory. In
order to reach this goal it is sufficient to study the uniqueness
of the solutions to the one-loop finiteness
conditions~\cite{KapMonZou93,LucZou97}. The Finite Unified ${\cal
N}=1$ supersymmetric SU(5) GUTs constructed in this way have
predicted correctly the top-quark mass, for example, from the
dimensionless sector (Gauge-Yukawa
unification)~\cite{KapMonZou93,KobKubMonZou01}. The search for RGI
relations has been extended to the soft supersymmetry breaking
sector (SSB) of these theories, which involves parameters of
dimensions one and two. In the SSB sector, besides the constraints
imposed by finiteness there are further restrictions imposed by
phenomenology. This in turn has led to a weakening of the
universality of soft scalar masses at the unification point and to
the introduction of a sum rule instead. In case the lightest
supersymmetric particle (LSP) is a neutralino the usual Higgs mass
is predicted to be in the range 115 - 130 GeV~\cite{LucZou96}.

Here we should like to continue the above point of view and consider
the singularities of general relativity as signaling a new structure
of space-time and in turn as a problem whose solution can eventually
be offered by noncommutative geometry. The ultimate aim is the
construction of a noncommutative generalization of the theory of
general relativity, which  becomes essentially noncommutative in
regions where the commutative limit would be singular. The physical
idea we have in mind is that the description of space-time using a
set of commuting coordinates is only valid at curvature scales
smaller than some fundamental one. At higher scales it is impossible
to localize a point and a new geometry should be used. We can think
of the ordinary Minkowski coordinates as macroscopic order
parameters obtained by ``course-graining" over regions whose size is
determined by a fundamental area scale $\kbar$, which is presumably,
but not necessarily, of the order of the Planck area $G \hbar$. They
break down and must be replaced by elements of a noncommutative
algebra when one considers phenomena at higher scales.

As a first concrete example we construct a modification of the
Kasner metric which is nonsingular whose singularity is resolved
into an essentially noncommutative structure. We recall that the
singularity of the Friedmann-Robertson-Walker isotropic
cosmological models, which constitute a basis for comparison of
theoretical predictions with observation is not sufficiently
general \cite{LanLif}. On the other hand the anisotropic Kasner
metric connected with the description of the oscillatory approach
to the cosmological singularity is considered sufficiently general
\cite{LanLif}. In addition we recall that Heckmann-Schucking type
of metrics \cite{KhaKam03} can bridge the Kasner with the
Friedmann-Robertson-Walker models which are suitable for
describing the later stages of the cosmological evolution.



It has been argued \cite{Mad00c} from simple examples that a
differential calculus over a noncommutative algebra uniquely
determines a gravitational field in the commutative limit. Some
examples have been given of metrics which resulted from a given
algebra and given differential calculus \cite{MacMad02,DimMad96}.
Here we aboard the inverse problem, that of constructing the
algebra and the differential calculus from the commutative metric.
As an example in which the above conjectures can be tested we
choose the Kasner-like metric which exhibits the most general
singularity.

In the following Greek indices take values from 0 to 3; the first
half of the alphabet is used to index (moving) frames and the second
half to index generators. Latin indices $a$, $b$, {\it etc.} take
values from 0 to $n-1$ and the indices $i$, $j$, {\it etc.} values
from 1 to 3. More details can be found in ref.~\cite{MacMad02}.

\initiate
\section{The general formalism}

%

According to the general idea outlined above a singularity in the
metric is due to the use of commuting coordinates beyond their
natural domain of definition into a region where they are
physically inappropriate. From this point of view the space-time
$V$ should be more properly described ``near the singularity" by a
noncommutative algebra $\c{A}$ over the complex numbers with four
hermitian generators $x^\lambda$.
We suppose also that there is a set of $n(=4)$ antihermitian
``momentum generators" $p_\alpha$ and a ``Fourier transform" $$ F:
x^\mu \longrightarrow p_\alpha = F_\alpha (x^\mu), $$ which takes
the position generators to the momentum generators.
If the metric which we introduce is the flat metric then we shall
see that $[p_\alpha, x^\mu] = \delta^\mu_\alpha$ and in this case
the ``Fourier transform" is the simple linear transformation
$$
p_\alpha = \frac{1}{i\kbar}\theta^{-1}_{\alpha\mu} x^\mu
$$
for some symplectic structure  $\theta^{\alpha\mu}$, that is an antisymmetric
non-degenerate matrix. As a measure of noncommmutativity, and to
recall the many parallelisms with quantum mechanics, we use the
symbol $\kbar$, which will designate the square of a real number
whose value could lie somewhere between the Planck length and the
proton radius. If the matrix is not invertible
then it is no longer evident that the algebra can be generated by
either the position generators or the momentum generators alone.
In such cases we define the algebra $\c{A}$ to be the one
generated by both sets.

We assume that $\c{A}$ has a commutative limit which
is an algebra $\c{C}(V)$ of smooth functions on a space-time $V$
endowed with a globally defined moving frame $\theta^\alpha$
which commutes with the elements of $\c{A}$, that is, for all $f
\in \c{A}$
\begin{equation}
f\theta^\alpha = \theta^\alpha f.                         \label{f-theta}
\end{equation}
We shall see that it implies that the metric components must be
constants, a condition usually imposed on a moving frame.
Following strictly what one does in ordinary geometry, we shall
chose the set of derivations
\begin{equation}
e_\alpha = \ad p_\alpha, \qquad e_\alpha f = [p_\alpha, f]
\label{GZ2}
\end{equation}
to be dual to the
frame $\theta^\alpha$, that is with
\begin{equation}
\theta^\alpha(e_\beta) = \delta^\alpha_\beta.\label{GZ3}
\end{equation}
We define the differential exactly as in the
commutative case. If $e_\alpha$ is a derivation of $\c{A}$ then
for every element $f \in \c{A}$ we define $df$ by the constraint
$df(e_\alpha) = e_\alpha f$.

It is evident that in the presence of curvature the 1-forms cease
to anticommute. On the other hand it is possible for flat ``space"
to be described by ``coordinates" which do not commute.  The
correspondence principle between the classical and noncommutative
geometries can be also described as the map
\begin{equation}
\t{\theta^\alpha}  \mapsto \theta^\alpha
\label{cp}
\end{equation}
with the product satisfying the condition
$$
 \t{\theta^\alpha}
\t{\theta^\beta} \mapsto P^{\alpha\beta}{}_{\gamma\delta}
\theta^\gamma \theta^\delta. $$
 The tilde on the left is to
indicate that it is the classical form. The condition can be
written also as
\begin{equation}
 \t{C}^\alpha{}_{\beta\gamma} \mapsto
C^\alpha{}_{\eta\zeta} P^{\eta\zeta}{}_{\beta\gamma}
\end{equation}
 or as
\begin{equation}\label{s-e1}
\lim_{\kbar\to 0}C^\alpha{}_{\beta\gamma} =
\t{C}^\alpha{}_{\beta\gamma}.
\end{equation}
We write $P^{\alpha\beta}{}_{\gamma\delta}$ in the form
\begin{equation}
P^{\alpha\beta}{}^{\phantom{\alpha\beta}}_{\gamma\delta} = \frac
12 \delta^{[\alpha}_\gamma \delta^{\beta]}_\delta + i\kbar\mu^2
Q^{\alpha\beta}{}^{\phantom{\alpha\beta}}_{\gamma\delta}.  \label{P}
\end{equation}
Flat noncommutative space is a solution to the problem of
constructing a noncommutative metric, given by the choice
\begin{equation}
e^\mu_\alpha = \delta^\mu_\alpha, \qquad K_{\alpha\beta} = - \frac
1{i\kbar} \theta^{-1}_{\alpha\beta} \in \c{Z}(\c{A}). \label{GZ}
\end{equation}
We have
introduced the inverse matrix $\theta^{-1}_{\alpha\beta}$ of
$\theta^{\alpha\beta}$; we must suppose the Poisson structure to
be non-degenerate: $\det \theta^{\alpha\beta} \neq 0$.  The
relations can be written in the form
\begin{eqnarray}
&&p_\alpha = - K_{\alpha\mu} x^\mu,\nonumber\\[4pt]
&&[p_\alpha, p_\beta] = K_{\alpha\beta}.         \label{flat}
\end{eqnarray}
This structure is flat according to our definitions. Here is
manifest one of the essential points of a Fourier transform. In
the limit when the $\theta_{\alpha\beta}$ tend to zero, the points
become well defined and in the opposite limit, when the
$\theta_{\alpha\beta}$ tend to infinity, the momenta become well
defined. In general Equation~(\ref{flat}) will be of the form
\begin{equation}
2 P^{\alpha\beta}{}_{\gamma\delta}p_\alpha p_\beta = K_{\gamma\delta}.         \label{curvedP}
\end{equation}
The corresponding rotation coefficients are given by
\begin{equation}
C^\alpha{}_{\gamma\delta} = -4 P^{\alpha\beta}{}_{\gamma\delta} p_\beta.         \label{curvedC}
\end{equation}

We shall find it convenient to consider a curved geometry as a
perturbation of a noncommutative flat geometry. The measure of
noncommutativity is the parameter $\kbar$; the measure of
curvature is a quantity $\mu^2$. We assume that $\kbar\mu^2$ is
small and that in the the flat-space limit we have commutation
relations of the form $$ [x^\mu, x^\nu] = i\kbar J^{\mu\nu},
\qquad J^{\mu\nu} = \theta^{\mu\nu} ( 1 + O(i\kbar \mu^2)). $$

\initiate
\section{The commutative Kasner metric}                      \label{classical}

A major problem
is the choice of an appropriate frame.
%
Given a symmetric matrix $Q = (Q^a_b)$ of real numbers, one possibility
for the Kasner metric is given by
\begin{equation}
\t{\theta}^0 = d\t{t}, \qquad \t{\theta}^a = d\t{x}^a - Q^a_b
\t{x}^b  \t{t}^{-1} d\t{t}.   \label{3.1}
\end{equation}
The 1-forms $\t{\theta}^\alpha$ are dual to the derivations $$
\t{e}_0 = \t{\partial}_0 + Q^i_j \t{x}^j
\t{t}^{-1}\t{\partial}_i, \qquad \t{e}_a = \t{\partial}_a $$ of
the algebra $\c{A}$.
The Lie-algebra structure of the derivations is given by the
commutation relations
\begin{equation}
[\t{e}_a, \t{e}_0] = \t{C}^b{}_{a0} \t{e}_b, \qquad [\t{e}_a,
\t{e}_b] = 0                                      \label{ccr}
\end{equation}
with $$ \t{C}^b{}_{a0} = Q^b_a \t{t}^{-1}. $$
We have written the
frame in coordinates which are adapted to the asymptotic
condition.  There is a second set which is also convenient, with
space coordinates $x^{\prime a}$ given in matrix notation by
\begin{equation}
x^{\prime a} = (t^{-Q} x)^a.                    \label{t-q}
\end{equation}
The frame can be then written, again in matrix notation, with
space components in the form
$$
\theta^a = (t^Q d (t^{-Q} x))^a = (t^Q d x^{\prime})^a.
$$

The expression for $\t{C}^b{}_{a0}$ contains no parameters with
dimension but it has the correct physical dimensions. Let $G_N$ be
Newton's constant and $\mu$ a mass such that $G_N \mu$ is a length
scale of cosmological order of magnitude. As a first guess we
would like to identify the length scale determined by $\kbar$ with
the Planck scale: $\hbar G_N \sim \kbar$ and so we have $\kbar
\sim 10^{-87} \mbox{sec}^2$ and since $\mu^{-1}$ is the age of the
universe we have $\mu \sim 10^{-17} \mbox{sec}^{-1}$. The
dimensionless quantity $\kbar\mu^2$ is given by $\kbar\mu^2 \sim
10^{-120}$.  In the Kasner case the role of $\mu$ is played by
$\t{t}^{-1}$ at a given epoch $\t{t}_0$.

We shall see below that the spectrum of the commutator of two
momenta is the sum of a constant term of order $\kbar^{-1}$ and a
``gravitational" term of order $\mu \t{t}^{-1} = \kbar^{-1} \times
(\kbar \mu) \t{t}^{-1}$. So the gravitational term in the units we
are using is relatively important for $\t{t} \lesssim \kbar \mu$.
The existence of the constant term implies that the gravitational
field is not to be identified with the noncommutativity {\it per
se} but rather with its variation in space and time.

The components of the curvature form are given by
\begin{eqnarray}
&&\t{\Omega}^a{}_0 = (Q^2 -Q)^a_b \t{t}^{-2} \t{\theta}^0
\t{\theta}^b,                  \label{3.6a}
\\[6pt]
&&\t{\Omega}^a{}_b = - \tfrac 12 Q^a_{[c} Q_{d]b}
 \t{t}^{-2} \t{\theta}^c \t{\theta}^d.                    \label{3.6b}
\end{eqnarray}
The curvature form is invariant under a uniform scaling of all
coordinates. The Riemann tensor has components $$ \t{R}^a{}_{0c0}
= (Q^2 - Q)^a_b \t{t}^{-2}, \qquad \t{R}^a{}_{bcd} = Q^a_{[c}
Q_{d]b} \t{t}^{-2}. $$ The vacuum field equations reduce to the
equations $$
\tr (Q) = 1, \qquad \tr (Q^2) = 1.                      \label{v-s}
$$ If $q_a$ are the eigenvalues of the matrix $Q^a_b$ there is a
1-parameter family of solutions given by
\begin{equation}
q_a =\frac{1}{1+\omega+\omega^2}
\left(1+\omega,\;\omega(1+\omega),\; -\omega\right).
\label{pa}
\end{equation}
The most interesting value is $\omega=1$ in which case $$ q_a =
\tfrac 13 (2,\;2,\;-1). $$ The curvature invariants are
proportional to $\t{t}^{-2}$; they are singular at $\t{t} = 0$ and
vanish as $\t{t} \rightarrow \infty$.

The values $q_a = c$ for the three parameters are also of
interest. The Einstein tensor is given by $$ \t{G}^0_{0} = - 3 c^2
\t{t}^{-2} , \qquad \t{G}^a_{b} = - c (3c - 2)\delta^a_b
\t{t}^{-2}. $$ For the value $c=2/3$ the space is a flat FRW with
a dust source given by $$ \t{T}_{00} = - \frac 1{8\pi
G_N}\t{G}_{00} = \frac 1{6\pi G_N}. $$ For $c = 1/3$ the space is
Einstein with a time-dependent cosmological ``constant".

\initiate
\section{The algebra of the noncommutative Kasner metric}

We are now in a position to write the algebra of the
noncommutative Kasner metric. From the structure of the frame we
obtain commutation relations between the position and momentum
generators. Using these and Jacobi identities we determine the
momentum-momentum or the position-position
commutation relations.

\subsection{The position-momentum relations}

From the correspondence with the commutative limit of frame it is easy to
see that the position-momentum commutation relations are
\begin{equation}
\begin{array}{ll}
[p_0, t] = 1, & [p_0, x^b] = Q^b_c \tau x^c,\\[4pt] [p_a, t] = 0,
&[p_a, x^b] = \delta^b_a.
\end{array}                                       \label{duality}
\end{equation}
Note that we have introduced the element $\tau$ of the subalgebra
of $\c{A}$ generated by $t$ which must tend to a constant multiple
of $t^{-1}$ in the commutative limit.
In fact the derivations defined by Equation~(\ref{GZ2}) satisfy
Equation ~(\ref{GZ3}).

\subsection{The momentum-momentum relations}

\begin{figure}[t]
\centering
\includegraphics[angle=0,scale=0.9]{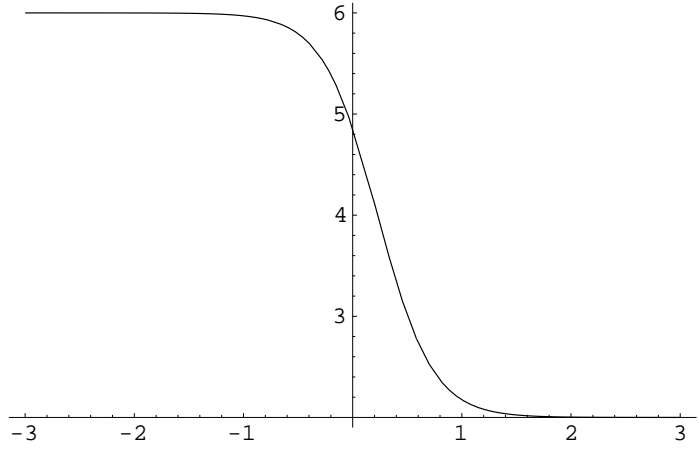}
\caption{A non-singular solution for $\tau$. (Horizontal axis t in
units of $(m\sqrt{|q|})^{-1}$, $\tau$ in units of $m\sqrt{|q|}$).}
\label{f1}
\end{figure}

\begin{figure}[t]
\centering
\includegraphics[angle=0,scale=0.9]{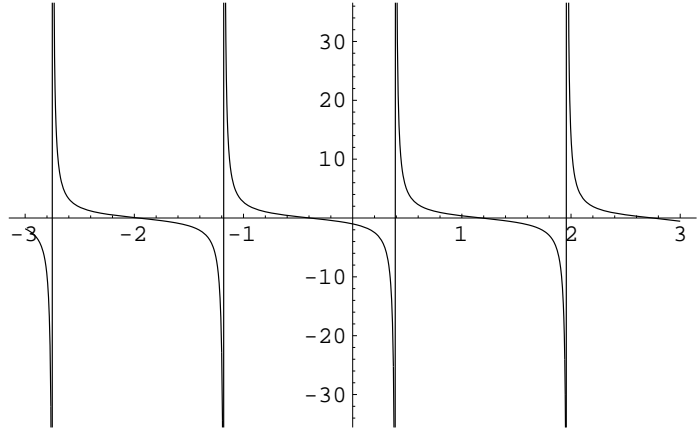}
\caption{A singular solution for $\tau$. (Horizontal axis t in
units of $(m\sqrt{|q|})^{-1}$, $\tau$ in units of $m\sqrt{|q|}$).}
\label{f2}
\end{figure}

We write the commutation relations~(\ref{curvedP}) satisfied by
the momentum generators $p_{\alpha}$ using Equation~(\ref{P}) and
the Ansatz
$$
Q^{cd}{}_{a0} = \tfrac 18 k^{(c} Q^{d)}_{a}, \qquad
Q^{cd}{}_{ab} = \tfrac 14 k^{c} k^{d} K_{ab}.
$$
We obtain then the relations
\begin{eqnarray}
\left[ p_{a}, p_{b} \right] = K_{ab} + L_{ab}(\tau), \label{first}\\
\left[ p_{0}, p_{a} \right] = K_{0a} + L_{0a}(\tau).  \label{second}
\end{eqnarray}
To be consistent with the
commutative limit the $L_{0a}$ must be given by
\begin{equation}
L_{0a}(\tau) =  Q_{a}^{b}\tau p_{b}.
\end{equation}
%
%
The element $\tau$ tends to a multiple of
$t^{-1}$ of the commutative algebra. We choose
$k^{a}$ to be an eigenvector of $Q_{a}^{b}$ with eigenvalue $q$,
that is $Q_{a}^{b} k^{a} = q k^{b}$. This choice simplifies the
calculations to be performed. To further simplify we choose $k^{a}
= (0,0,1)$, $K_{0a} = - \frac{1}{i \kbar} l_{a}$, $l_{a} = (0, 0,
l)$. Then Equation~(\ref{second}) can be written as
\begin{eqnarray}
\left[ p_{0}, p_{3} \right] &=&K_{03} - q \tau p_{3}= -\frac{1}{i \kbar} l, \label{explicit}\\
\left[ p_{0}, p_{1} \right] &=& K_{01}=0, \nonumber \\
\left[ p_{0}, p_{2} \right] &=& K_{02}=0. \nonumber
\end{eqnarray}
The space derivations of the element $\tau$ must vanish; that is
$[p_{a}, \tau]=0$. To find an explicit form of $\tau$ we multiply
Equation~(\ref{explicit}) by $-i \kbar \mu^{2} q $ (where
$\mu^{2}$ is a scale of curvature as explained in section (2))
then we see that by introducing $\tau = -i \kbar \mu^{2} q p_{3} $
and setting $m^{2} = - i \kbar \mu^{2}q K_{03}$ we determine an
element of the algebra that has the property $[p_{a}, \tau] =0$
and obeys the differential equation
\begin{equation}\label{dynamical}
\dot{\tau} - q \tau^{2} + m^{2} = 0, \ \ \ m^{2} = \mu^{2} q
l.
\end{equation}
Let us first note an interesting duality of Equation
(\ref{dynamical}), namely
\begin{equation}\label{dudynamical}
\tau \rightarrow 1/\tau, \ \ m^{2} \rightarrow q.
\end{equation}
The generic solution of this equation has the form
\begin{equation}\label{sdynamical}
\tau = -\frac{1}{q} \frac{c_{1} |m| \sqrt{q}e^{+|m|\sqrt{q}t} +
c_{2} ( - |m| \sqrt{q})e^{-|m|\sqrt{q}t}}{c_{1}e^{+|m|\sqrt{q}t} +
c_{2}e^{-|m|\sqrt{q}t}},
\end{equation}
which e.g. when $c_{1} = c_{2}$  becomes
\begin{equation}
\tau = - \frac{1}{\sqrt{q}} |m| \tanh (\sqrt{q} |m| t).
\label{sol31}
\end{equation}
A general class of solutions of Equation~(\ref{sdynamical}) are
non-singular. A representative example is given in fig.~\ref{f1}
Depending on the value of $q$ we may obtain another general
class of periodic but singular solutions (like $\tau \sim \cot(t)$
) and a representative example is drawn in fig.~\ref{f2}.
The function $\tau$ enters the curvature invariants
\cite{MacMad02} and in case it is smooth, the singularity problem
is avoided.

From the Jacobi identities we find that $L_{ab}$ must be of the
form
\begin{equation}
L_{ab} = \frac{q}{m^{2}}K_{ab}\tau^{2}
\end{equation}
as well as the algebraic condition $$ 2q K_{ab} =
Q^{c}_{[a}K_{b]c},
$$ which reduces to $$ TrQ = -3q, \ \ q_{1} + q_{2} + q = - 3q.$$
In the following we think of $Q$ as the matrix $Q= diag(q_{1},
q_{2}, q)$. We have determined the lowest order correction to
$L_{ab}$, while the series does not seem to sum to a simple known
function.

The momentum algebra of Equation~(\ref{P}) takes in the present
case the form
\begin{eqnarray}
&& [p_0, p_a] = K_{0a} +  Q^b_a  \tau p_b, \label{ll1}
\\[4pt]
&& [p_a, p_b] =   K_{ab}(1+
\frac{q\tau^{2}}{\mu^{2}}). \label{ll2}
\end{eqnarray}
The above form of the momentum algebra can be derived from the
following choice for $Q^{ab}_{\ \ cd}$ and $Q^{ab}_{\ \ 0c}$,  $$
Q^{ab}_{\ \ cd} = -i \kbar \frac{m^{2}}{2ql}
k^{a}k^{b}K_{cd}$$ and $$Q^{ab}_{\ \ 0c}=
\frac{m^{2}}{2l} Q^{a}_{c} k^{c}.$$ The commutation relations
are then
\begin{eqnarray}
&& [p_a, p_b] = (i\kbar)^{-1} \epsilon_{abc}k^c + \textstyle{\frac
12} p_c C^c{}_{ab} \nonumber
\\[4pt]
&& \phantom{[p_a, p_b]} = (i\kbar)^{-1} \epsilon_{abc}k^c - 2
i\kbar p_c p_d Q^{cd}{}_{ab}               \nonumber
\\[4pt]
&& \phantom{[p_a, p_b]} = (i\kbar)^{-1} \epsilon_{abc}k^c(1+
 \frac{q \tau^{2}}{m^{2}}), \label{con01}
\\[8pt]
&& [p_0, p_a] = (i\kbar)^{-1} l_a + \textstyle{\frac 12} p_c
C^c{}_{0a}                               \nonumber
\\[4pt]
&& \phantom{[p_0, p_a]} = (i\kbar)^{-1} l_a - 2i\kbar p_b p_c
Q^{bc}{}_{0a}                      \nonumber
\\[4pt]
&& \phantom{[p_0, p_a]} = (i\kbar)^{-1}l_a +\tau Q^d_a p_d.
                                                               \label{con02}
\end{eqnarray}

To simplify the commutation relations we have chosen $k^a = (0,0,
1)$ and $l_{a} = (0,0,l) $,  then we have
\begin{equation}
\begin{array}{ll}
[p_{2}, p_{3}] = 0, & [p_{3}, p_{1}]=0,
\\ [4pt]
[p_1, p_2] = (i \kbar)^{-1}(1 + \frac{q \tau^{2}}{m^{2}}), & [p_0,
p_1]  =   q_1 p_1 \tau,
\\[4pt]
[p_0, p_2]  =  q_2 p_2 \tau, & [p_0, p_3] = (i\kbar)^{-1} l + q
p_{3} \tau.
\end{array}                                          \label{adapted}
\end{equation}
It is possible to represent $\c{A}$ as a tensor product of two
Heisenberg algebras, i.e. $$ \c{A}_{12} \otimes \c{A}_{30} \subset
\c{A} $$ We denote the generators of $\c{A}_{12}$ by $\mu_{1},
\mu_{2}$ and the generators of $\c{A}_{30}$ by $\mu_{0}, \mu_{3}$.
Being in the system $(0, 0, 1)$ the relation between $\tau$ and
$p_3$ becomes
\begin{equation}
p_{3} = - \frac{1}{i \kbar \mu^{2} q } \tau,
\end{equation}
e.g. using Equation~(\ref{sol31}) we obtain
\begin{equation}
p_{3} = \frac{ |m| }{ i \kbar \mu^{2} q^{3/2} } \tanh(|m|\sqrt{q}
t).
\end{equation}
Then we define $\mu_{0} = p_{0}$ and $\mu_{3} = \frac{1}{i \kbar}
t $ and we define $\mu_{1}$ and $\mu_{2}$ in such a way that
\begin{eqnarray}
\left[ \mu_{0}, \mu_{1} \right] = 0, & & \left[ \mu_{3}, \mu_{1}
\right] = 0, \nonumber
\\ \left[ \mu_{0},\mu_{2} \right] = 0,  & & \left[ \mu_{3}, \mu_{2}
\right] = 0.
\end{eqnarray}
But
\begin{equation}
[\mu_{0}, \mu_{3}] \neq 0,\ \ \ [\mu_{1}, \mu_{2}] \neq 0.
\end{equation}
To achieve this we set $p_{1} = \mu_{1}U_{1}, p_{2} =
\mu_{2}U_{2}$ where $U_{1,2} = U_{1,2}(t)$ and calculate the
commutation relations $[\mu_{0} , p_{1,2}]$. We find that the
$U$'s must satisfy the equation
\begin{equation}\label{Q}
\dot{U} = Q \tau U
\end{equation}
in order our commutation relations for $\mu_{\alpha}$ to have the
desired form. We denote a solution of Equation~(\ref{Q}) depending
on the parameter $Q$ as $U_{(Q)}$, i.e. $U_1 \equiv U_{(q_{1})},
U_2 \equiv U_{(q_{2})}$. Next calculating $[p_{1}, p_{2}]$ we find
that
\begin{equation}
[\mu_{1}, \mu_{2}] = \frac{1}{i \kbar U_{1} U_{2}} (1+
 \frac{q\tau^{2}}{m^{2}}).
\end{equation}
Similarly calculating $[\mu_{0}, p_{3}] $ we find
\begin{equation}
[\mu_{0}, \mu_{3}] = \frac{1}{i \kbar}.
\end{equation}

\initiate
\subsection{The position-position relations}

In the simplified  momentum basis we make an ansatz for the
Fourier transform relating the coordinates $x,y,z,t$ to momenta
$\mu_1, \mu_2, \mu_0, \mu_3$
\begin{eqnarray}\label{Position}
x&=& i \kbar \mu_{1} f_{1}, \nonumber\\ y&=& i \kbar
\mu_2 f_{2}, \nonumber\\ z&=& i \kbar \mu_{0}f_{3},\nonumber\\
t&=& i \kbar \mu_3,
\end{eqnarray}
where $f_{1,2,3}$ are functions of $t$ to be determined from the
momentum-position commutation relations. Putting
eqs.~(\ref{Position}) to eqs.~(\ref{duality}) we find that $$
f_{1} = U_{(q_{1})},$$ $$f_{2} = U_{(q_{2})},$$ $$ f_{3} =
U_{(q)}, $$ where the functions $U$ were defined above and depend
on the parameter indicated in the subscript. We can now calculate
the commutation relations between the coordinates and find
\begin{equation}
\begin{array}{ll}
[x,y] = i \kbar \rho, & [x,z] = -i \kbar q_{1} \tau x U_{(q)}
\\ [4pt] [y,z] = -i \kbar q_{2} \tau y U_{(q)}, & [t,z] = i \kbar U_{(q)}.
\end{array}
\end{equation}
where
$$
\rho = (1 +  \frac{q \tau^{2}}{m^{2}}).
$$
Then the $J^{\mu\nu}$ tensor is given by
 \begin{eqnarray*}
J^{\mu\nu} = U_{(q)}^{-1} \left(\begin{array}{cccc} 0 & \rho
U_{(q)} & -q_1 \tau x & 0\\[4pt] -\rho U_{(q)} & 0 &  -q_2 \tau y
& 0\\[4pt] q_1 \tau x & q_2 \tau y & 0 & -1 \\[4pt] 0  & 0 & 1 & 0
\end{array}\right).
\end{eqnarray*}
This can be written in the form
$$
J^{\mu\nu} = S^{\mu\nu} + x^{[\mu} P^{\nu]}
$$
with
 \begin{eqnarray*}
S^{\mu\nu} = \left(\begin{array}{cccc} 0 & \rho & 0 & 0\\[4pt] -\rho  & 0 &  0
& 0\\[4pt] 0 & 0& 0 & -1 \\[4pt] 0  & 0 & 1 & 0
\end{array}\right),
\end{eqnarray*}
and
$$
P^\mu = U_{(q)}^{-1}\tau(0, \,0 , \,0, \, -q_1).
$$
We see then from the behaviour of $\tau$ that the
commutation relations diverge at the origin. Indeed the orbital
term $x^{[\mu} P^{\nu]}$ vanishes exponentially and the spin term
diverges as $t^{-2}$.

%
 $$
 $$
\initiate
\section{Conclusions}                            \label{cosm}
In conclusion we propose to resolve the singularities of general
relativity by assuming that space-time becomes fuzzy beyond a
certain scale. In the specific example we have given here  the
Kasner manifold has been replaced by a noncommutative algebra,
whose Jacobi identities force a modification of the time
dependence of the metric. All curvature invariants depend smoothly
on a element $\tau$, which replaces the time coordinate. We have
seen that particular choices of parameters in
Equation~(\ref{sdynamical}) lead to nonsingular solution such as
Equation~(\ref{sol31}) which is a desirable result for the
programme we have put forward.

We note that the above nonsingular solution has the interesting
property of extrapolating between two flat solutions of different
(constant) commutation relations, that is in the notation of
section 2 $$ \theta_{-}^{\alpha \beta} = \lim_{t \rightarrow -
\infty} J^{\alpha \beta}$$ and $$\theta_{+}^{\alpha \beta} =
\lim_{t \rightarrow + \infty} J^{\alpha \beta}$$ are not equal and
can be arbitrary. In particular one of them can vanish. In this
way we have a smooth extrapolation between a noncommutative flat
space and a commutative one. On the other hand the general
solution given by Equation~(\ref{sdynamical}) (depending on the
value of $q$) contains periodic solutions which are singular. The
duality (\ref{dudynamical}) of Equation (\ref{dynamical}) connects
the singular points with the regular points.

It should be stressed though that no use is made of field
equations. The restrictions on the solutions find their origin in
the requirement that the noncommutative algebra be an associative
one and appear as Jacobi identities on the commutation relations.

\initiate
\section*{Acknowledgments} One of the authors (JM) would like to thank
M.~Buri\'{c}, S.~Cho, G.~Fiore, E.~Floratos, H.~Grosse and P.~Tod
for interesting conversations. Preliminary work was to a large
extent carried out while he was visiting the MPI in M\"unchen. He
is grateful to Julius Wess also for financial support during this
period. (JM), (PM) and (GZ) would like to thank A.~Dimakis for
useful discussions. They acknowledge partial support by EU under
the RTN contract HPRN-CT-2000-00148, the Greek-German Bilateral
Programme IKYDA-2001 and by  the NTUA programme for fundamental
research ``THALES". (PM) is supported by a ``PYTHAGORAS"
postdoctoral fellowship.


\providecommand{\href}[2]{#2}\begingroup\raggedright\endgroup

\end{document}